\documentclass[a4paper,fleqn,useAMS,usenatbib]{mnras}

\pdfoutput=1

\usepackage{mathptmx}

\usepackage[T1]{fontenc}
\usepackage{ae,aecompl}

\usepackage[dvips]{graphicx}
\usepackage{amsmath}
\usepackage{amsfonts}
\usepackage{amssymb}
\usepackage{comment}
\usepackage[dvipsnames]{xcolor}
\usepackage[%
        ]{hyperref}
        
\hypersetup{
	colorlinks=true,	
	urlcolor=MidnightBlue,		
	pdfpagelabels=true,
	hypertexnames=true,
	plainpages=false,
	naturalnames=true,
	pdftitle={A Cloaking Device for Transiting Planets},    
	pdfauthor={Kipping \& Teachey},     
	linkcolor=WildStrawberry,          
	citecolor=ForestGreen,        
}

\newcommand{\KeplerMission}{ \emph{Kepler Mission} }

\newcommand{\Wrec}{W_{\mathrm{r}}}
\newcommand{\Wtra}{W_{\mathrm{t}}}

\newcommand{\Drec}{D_{\mathrm{r}}}
\newcommand{\Dtra}{D_{\mathrm{t}}}

\newcommand{\Prec}{\mathcal{P}_{\mathrm{r}}}
\newcommand{\Ptra}{\mathcal{P}_{\mathrm{t}}}
\newcommand{\Etra}{\mathcal{E}_{\mathrm{t}}}

\newcommand{\Precstar}{\mathcal{P}_{\star,\mathrm{r}}}
\newcommand{\Ptrastar}{\mathcal{P}_{\star,\mathrm{t}}}

\newcommand{\Kepler}{ {\it Kepler} }

\title[A Cloaking Device for Transiting Planets]{A Cloaking Device for Transiting Planets}
\author[Kipping \& Teachey]{David M. Kipping$^{1}$\thanks{E-mail:
\href{mailto:dkipping@astro.columbia.edu}{dkipping@astro.columbia.edu}} and Alex Teachey$^{1}$\\
$^{1}$Dept. of Astronomy, Columbia University, 550 W 120th Street, New York NY 10027}

\date{Accepted . Received ; in original form }

\pubyear{2016}

\begin{document}
\label{firstpage}
\pagerange{\pageref{firstpage}--\pageref{lastpage}}
\maketitle

\begin{abstract}
	
The transit method is presently the most successful planet discovery and characterization
tool at our disposal. Other advanced civilizations would surely be aware of this technique
and appreciate that their home planet's existence and habitability is essentially broadcast
to all stars lying along their ecliptic plane. We suggest that advanced civilizations could
cloak their presence, or deliberately broadcast it, through controlled laser emission. Such
emission could distort the apparent shape of their transit light curves with relatively 
little energy, due to the collimated beam and relatively infrequent nature of transits. We 
estimate that humanity could cloak the Earth from \Kepler-like broadband surveys using an 
optical monochromatic laser array emitting a peak power of $\sim$30\,MW for $\sim10$\,hours 
per year. A chromatic cloak, effective at all wavelengths, is more challenging requiring a 
large array of tunable lasers with a total power of $\sim$250\,MW. Alternatively, a 
civilization could cloak only the atmospheric signatures associated with biological activity 
on their world, such as oxygen, which is achievable with a peak laser power of just 
$\sim$160\,kW per transit. Finally, we suggest that the time of transit for optical SETI is 
analogous to the water-hole in radio SETI, providing a clear window in which observers may 
expect to communicate. Accordingly, we propose that a civilization may deliberately broadcast 
their technological capabilities by distorting their transit to an artificial shape, which 
serves as both a SETI beacon and a medium for data transmission. Such signatures could be 
readily searched in the archival data of transit surveys.

\end{abstract}


\section{Introduction}
\label{sec:intro}

In the last two decades, a new era of astronomical discovery has blossomed through
the detection of planets orbiting other stars, exoplanets. In the first decade,
this wave of discovery was largely enabled by precise measurements of the reflex
radial velocity of stars, apparent via the spectroscopic shifts of atomic
absorption lines \citep{mayor:1995,marcy:1996}. From 1995--2005, 95\% of
exoplanet discoveries were achieved using radial velocities
(\href{http://www.exoplanets.org}{Exoplanet Orbit Database}; \citealt{wright:2011}).
During this time, only a handful of planets were found via the photometric
brightness decrease of a star caused by a transiting planet
\citep{charbonneau:2000,henry:2000}. However, once demonstrated, the relative
technical ease, potential sensitivity and wide scope for surveying led to 
the transit method to become the dominant method of finding other planets.
From 2005--2015, 75\% of exoplanet discoveries were made using transits,
of which over 80\% were from the \KeplerMission\ \citep{borucki:2003}.
With multiple transit surveys planned over the next decade, such as NGTS
\citep{wheatley:2013}, TESS \citep{ricker:2014} and PLATO \citep{rauer:2014},
this pattern of discovery seems likely to continue.

One of the major goals of these exoplanet surveys is to improve our understanding
regarding our uniqueness in the Universe, to resolve whether the Copernican
Principle \citep{bondi:1952} extends to self-aware, technological civilizations.
Whilst this is usually framed as measuring the occurrence rate of Earth-like
planets (see, e.g. \citealt{petigura:2013}), a key term in the
Drake Equation, transits can go further and more directly answer this most
fundamental of questions. For example, transit spectroscopy \citep{seager:2000}
searches for the chromatic transit variations due to absorbing molecules. 
So-called ``biosignatures'', such as molecular oxygen \citep{desmarais:2002,
kaltenegger:2002}, could be revealed this way, although abiotic activity could be 
ultimately responsible too \citep{wordsworth:2014}. \citet{lin:2014} proposed that
advanced civilizations could be identified by detecting the atmospheric pollutants
produced by industrial activity, although it is doubtful that present anthropogenic
activity is sustainable \citep{field:2014}. These ambiguities would be eliminated 
if an advanced civilization chose to deliberately signal their presence or undertake 
mega-engineering in their system. Here also, the success of recent transit surveys 
has motivated thinking as to whether transits could serve such a function too.

Transiting mega-structures, such as Dyson spheres \citep{dyson:1960}, would present
irregular transit profiles which may be difficult to distinguish from natural phenomena,
such as a cometary clouds \citep{wright:2016,boyajian:2016}. Deliberate signaling
(or broadcasting) would be presumably easier to interpret, which motivated the original
radio-based ``Search for Extraterrestrial Intelligence'' (SETI) project \citep{morrison:1977}.
\citet{arnold:2005} proposed that advanced civilizations could broadcast
their presence through transits by constructing large, thin artificial structures,
such as a triangular mask, in a tight orbit around their host star. The transit
profiles of such objects would display residuals to the assumed spherical planet model,
on the order of $10^{-4}$. A related concept using mirrors was recently proposed by
\citet{korpela:2015}, but both ideas require the construction of Earth-sized masks,
which is far beyond our current capabilities.

In this work, we argue that artificial transit profiles can be feasibly generated
using laser emission. Whilst typical optical SETI efforts focus on detecting brief
pulse-like lasers at specific wavelengths (see, e.g. \citealt{tellis:2015}), we
propose that deliberate distortion of one's own transit profile can be an effective
means of signaling too. Moreover, an advanced civilization could plausibly cloak
their planet (or moon) from \Kepler-like transit surveys at relatively low energetic
cost. Indeed, it is technologically feasible that we could soon begin
cloaking the Earth from nearby Earth-analogs lying near the ecliptic
plane (such that they see us transit) with this technique.

Starting with the cloaking case, which is conceptually simpler, we derive the energy
requirements to cloak a transiting planet in \S\ref{sec:cloak}, both in terms of
individual events and the continuous power required averaged over the planet's
orbital period. We build upon this idea in \S\ref{sec:biocloak}, introducing the 
more energetically favorable case of a ``biocloak'', a cloak of biosignatures alone.
In \S\ref{sec:broadcast}, we propose a possible method for broadcasting
via lasers to create an unambiguously artificial transit revealing our presence. We
conclude in \S\ref{sec:discussion} discussing the possibilities to actively search for
artificial transits in archival and upcoming photometric mission data.

\section{Cloaking A Planet}
\label{sec:cloak}

\subsection{Concept}
\label{sub:cloakconcept}

Consider an advanced civilization residing on a planet\footnote{Whilst we use the
term ``planet'' throughout, this could equally apply for moons or trojans too.},
which would presumably have similar insolation to that of the Earth, $S_{\oplus}$.
Let us imagine that the inhabitants have discovered all the nearby habitable
planets along their ecliptic plane. Since the inhabitants presumably know the
orbit of their own planet precisely and the positions of the stars hosting
habitable planets, they are easily able to evaluate which habitable worlds would 
be able to observe transits of their home planet and when.

Given humanity's own experience of highly successful transit surveys as a means of
discovering exoplanets, it is not unreasonable that many extraterrestrial 
civilizations may have adopted this method as part of their planet hunting arsenal.
With this realization, the hypothetical advanced civilization may wish to avoid other
civilizations discovering their presence. Since this civilization knows precisely when
other nearby habitable planets could see them transit, their objective would be to
generate a brightness change, as viewed by a distant observer, which would cancel out
the transit light curve caused by their home planet eclipsing their parent star.

One method to achieve this requirement would be with a fleet of finely controlled
mirrors, analogous to the mega-engineering project discussed by \citet{korpela:2015}.
The dimensions, number and geometry of these mirrors would require carefully tuning
for each target, at considerable expense to build and launch these structures.
Whilst this elaborate method could work, it is arguably technologically more feasible,
given our present state of development, to use a directed laser beam to achieve the same
effect.

\begin{figure*}
\begin{center}
\includegraphics[width=16.8 cm]{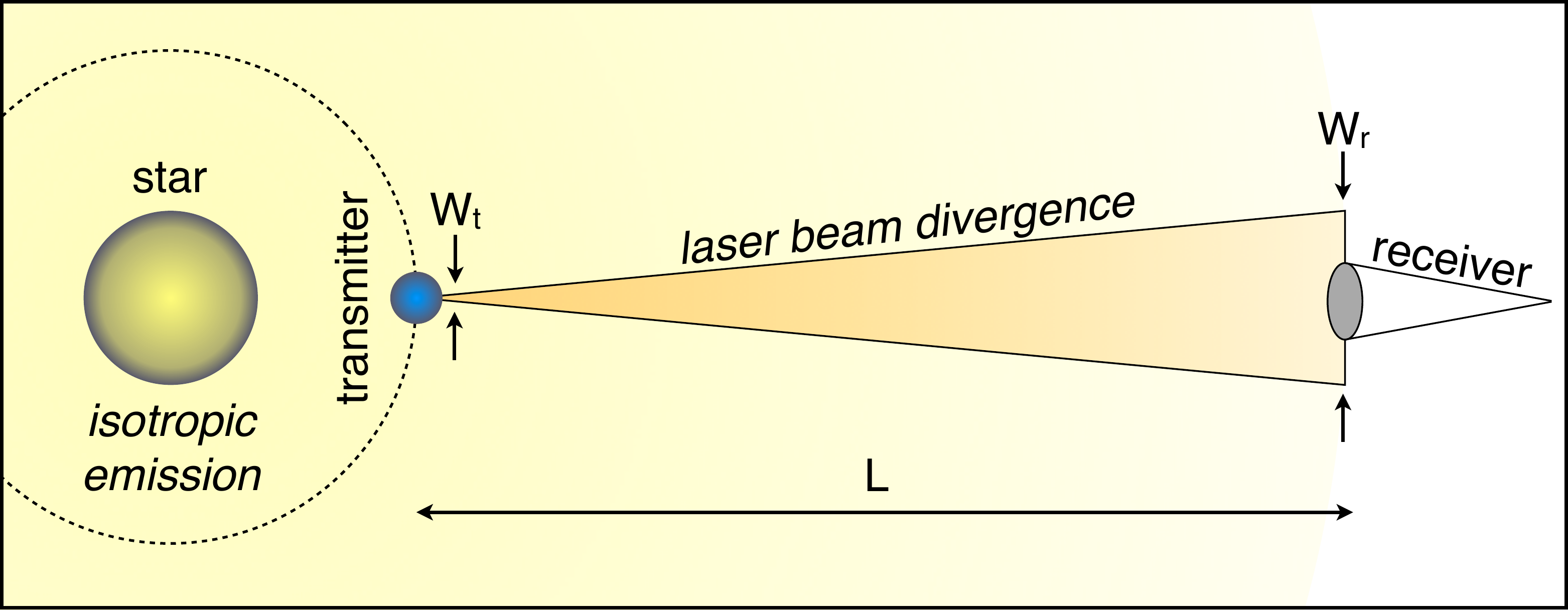}
\caption{
Illustration (not to scale) of the transit cloaking/broadcasting device. A laser beam
(orange) is fired from the night side of inhabited planet (blue) towards a target star
whilst the planet appears to transit the star, as seen from the receiver. In the case
of the Earth, the planet could be cloaked by generating an inverse transit-like signal
of peak power $\lesssim$60\,MW.
}
\label{fig:cartoon}
\end{center}
\end{figure*}

In our scheme, as depicted in Figure~\ref{fig:cartoon}, the advanced civilization
emits a laser directed towards the other planetary system at precisely the instant
when the other system would be able to observe a transit. The power profile of the
laser would need to be the inverse of the expected transit profile, leading to a
nullified flat line eliminating the transit signature.

\subsection{Broadband Cloak}
\label{sub:powerreq}

In what follows, we consider the power and energy requirements needed to produce an
inverted transit signal. For simplicity, we will first assume that the observer we wish
to cloak from is using a \Kepler-like survey, defined as an optical, broadband
photometer, and thus we refer to this as a ``broadband'' cloaking device. The issue of
cloaking transits from spectroscopic transit measurements
is discussed later in \S\ref{sub:chromatic}.

In the described scenario, one can utilize a monochromatic optical laser, which would
emit at a wavelength around the broad high efficiency plateau of an optical survey
(e.g. for \Kepler\, the response function plateaus for 450-800\,nm). For simplicity,
we will assume a bolometric bandpass, collecting photons of all wavelengths, in what
follows. Modifying the final result for narrower bandpasses is easily achieved,
as demonstrated towards the end of our calculation.

Consider a laser with a beam width at the transmitter and receiver of $\Wtra$
and $\Wrec$ respectively, where the separation between them is given by $L$,
as depicted in Figure~\ref{fig:cartoon}. The beam divergence angle, $\Theta$,
is defined by

\begin{align}
\Theta = 2\tan^{-1} \Big( \frac{ \Wrec - \Wtra }{ 2 L } \Big).
\end{align}

\noindent One may re-arrange this to express that

\begin{align}
\Wrec &= \Wtra + 2 L \tan(\Theta/2).
\end{align}

Consider the case of $L$ being very large, such that significant divergence will occur,
causing $\Wtra \ll \Wrec$ and thus the above can be reasonably approximated to

\begin{align}
\Wrec &= 2 L \tan(\Theta/2),
\label{eqn:Wrec}
\end{align}

\noindent For a diffraction limited beam, the divergence angle is given by

\begin{align}
\Theta &= \frac{ \sqrt{2} \lambda}{ \Dtra },
\end{align}

\noindent which may be substituted into Equation~(\ref{eqn:Wrec}) to yield

\begin{align}
\Wrec &= \frac{ \sqrt{2} L \lambda }{ \Dtra },
\label{eqn:Wrec2}
\end{align}

\noindent where we have used the small-angle approximation for $\Theta$. The receiver has
a finite diameter, $\Drec$, which we assume to be less than the diameter of the
beam width at the receiving end (i.e. $\Drec \leq \Wrec$). In canonicalized units,
the beam width at the receiver is

\begin{align}
\Wrec &\simeq 0.1\,\mathrm{AU} \Big(\frac{L}{10\,\mathrm{pc}}\Big) \Big(\frac{\lambda}{600\,\mathrm{nm}}\Big) \Big(\frac{10\,\mathrm{m}}{\Dtra}\Big).
\label{eqn:Wrec3}
\end{align}

This suggests that for systems closer than 100\,pc, the beam width will not
exceed an AU and some information about the target planet's orbit would be
required. The pointing precision is expected to be given by 
$\alpha=\tfrac{1}{2} \Wrec/L$, or

\begin{align}
\alpha &\simeq 10\,\mathrm{mas}\, \Big(\frac{\lambda}{600\,\mathrm{nm}}\Big) \Big(\frac{10\,\mathrm{m}}{\Dtra}\Big).
\label{eqn:pointing}
\end{align}

By placing the receiver within the beam cone, the fraction of 
transmitted power collected by the receiver will be equal to the ratio-of-areas 
between the receiver and the beam width\footnote{In practice, the intensity of a
standard laser would be non-uniform across the beam, but this may be correctable
using beam shaping, as discussed later in \S\ref{sub:laserreq}.}, such that

\begin{align}
\Prec &= \Bigg(\frac{ \Drec }{ \Wrec }\Bigg)^2 \Ptra,
\label{eqn:powerassumption}
\end{align}

\noindent where $\Ptra$ and $\Prec$ is the power transmitted/received by the transmitter/receiver,
respectively. Substituting in Equation~(\ref{eqn:Wrec2}) for $W_{\mathrm{r}}$, yields

\begin{align}
\Prec &= \Bigg(\frac{ \Drec \Dtra }{ \sqrt{2} L \lambda }\Bigg)^2 \Ptra.
\label{eqn:Prec}
\end{align}

The parent star of this planetary system is also a power source, producing a
power of $\Ptrastar$. In the case of a bolometric receiver, the power seen
to be emitted from the source equals the stellar luminosity, $\Ptrastar = L_{\star}$.
If the star and laser are unresolved sources on the sky, then the receiver will
collect photons from both sources, with the component coming from the star being
equal to

\begin{align}
\Precstar &= \Ptrastar \Bigg( \frac{ \pi \Drec^2 }{ 4\pi L^2 } \Bigg).
\label{eqn:Pstarrec}
\end{align}

In order to cloak a transiting planet, we require the power received from
the laser to nullify that of the transit signal. The maximal laser power
required would occur at the time of transit minimum (defined as the instant
when the sky-projected planet-star separation is minimized), at which
point the power balance would need to satisfy $\Prec \simeq (R_P/R_{\star})^2
\Precstar$, where $(R_P/R_{\star})$ is the ratio-of-radii between the planet
selected for cloaking and the parent star. Substituting Equation~(\ref{eqn:Prec})
for $\Prec$ and Equation~(\ref{eqn:Pstarrec}) for $\Precstar$, this condition
implies

\begin{align}
\mathcal{P}_{\mathrm{t}} &= \Bigg( \frac{ R_P^2 \lambda^2 }{ 2 D_{\mathrm{t}}^2 R_{\star}^2 } \Bigg) \mathcal{P}_{\star,\mathrm{t}}.
\label{eqn:Pt}
\end{align}

\noindent Using the Stefan-Boltzmann Law for the stellar luminosity, allows us to
simplify the above to

\begin{align}
\mathcal{P}_{\mathrm{t}} &= \frac{ 2 \pi \sigma_B R_P^2 \lambda^2 T_{\star}^4 }{ D_{\mathrm{t}}^2 }.
\end{align}

\noindent Assuming that most of the power is produced near $\lambda=600$\,nm, the
peak wavelength of Sun-like stars, the approximate laser power requirement
is

\begin{align} 
\mathcal{P}_{\mathrm{t}} &= 58\,\mathrm{MW}\,\Big( \frac{R_P}{R_{\oplus}} \Big)^2 \Big(\frac{\lambda}{600\,\mathrm{nm}}\Big)^2 \Big( \frac{ D_{\mathrm{t}} }{ 10\,\mathrm{m} } \Big)^{-2} \Big( \frac{ T_{\star} }{ 5777\,\mathrm{K} } \Big)^4.
\label{eqn:physicalpower}
\end{align}

The actual peak power would depend on the degree to which limb darkening 
affects the light curve, leading to a modest increase in the power requirement
near the center of the transit and a decrease near the limb. For Solar-like
limb darkening in the \Kepler\ bandpass, for example, this equates to a
$\sim20$\% increase in peak power.

\begin{figure}
\begin{center}
\includegraphics[width=8.4 cm]{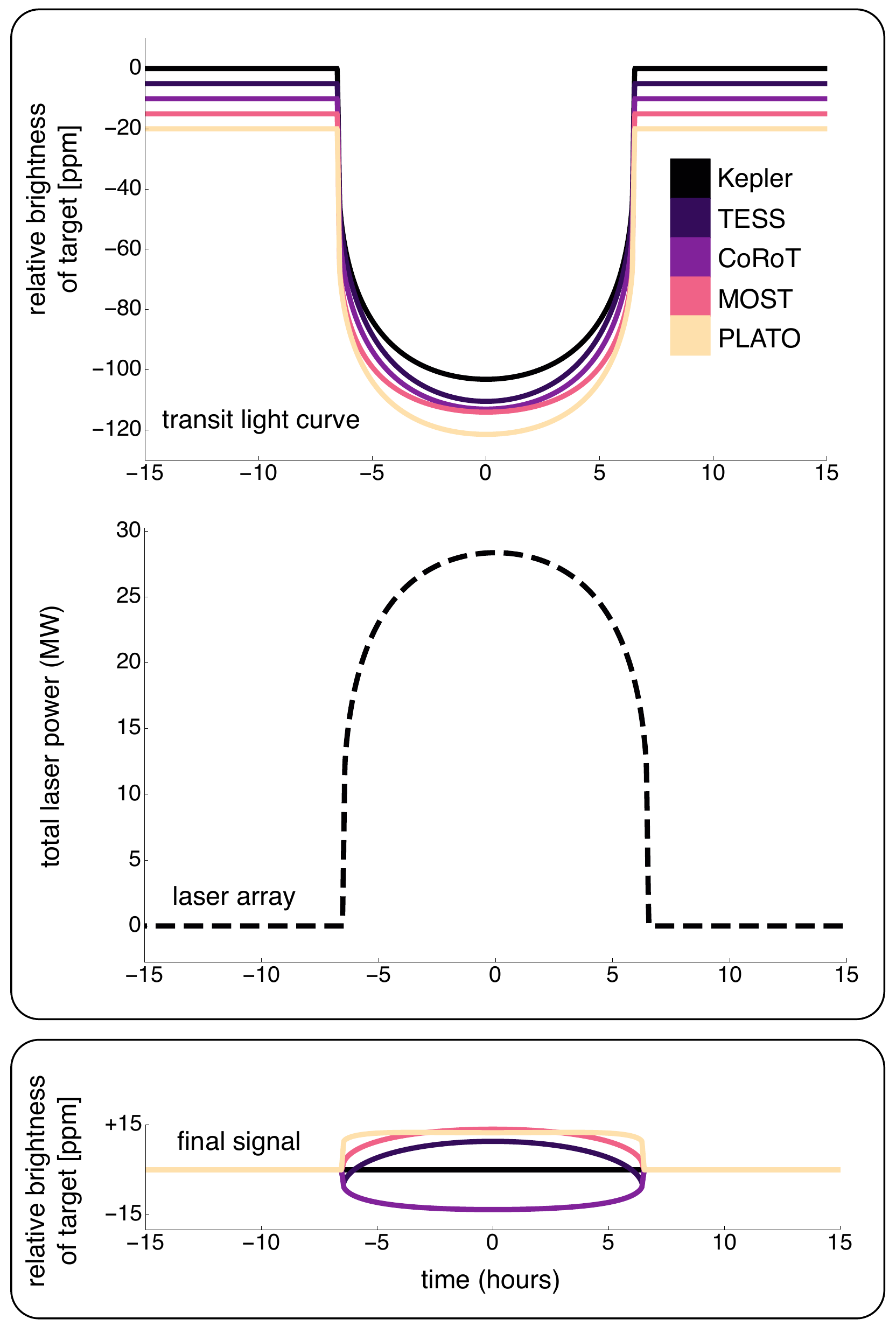}
\caption{
\textit{Top:} The unaltered light curve of the Earth transiting the Sun,
as viewed by different broadband optical photometers (offset by 5\,ppm).
\textit{Middle:} The power profile of a 600\,nm laser array designed to
cloak the Earth. An array of lasers producing a peak power of $\sim30$\,MW
over 13\,hours nullifies the transit.
\textit{Bottom:} Residual light curve, as seen by the different photometers.
}
\label{fig:cloak}
\end{center}
\end{figure}

Equation~(\ref{eqn:physicalpower}) reveals that the peak power required is
$\sim60$\,MW, although cloaking a planet from an observer using a narrower
bandpass than bolometric would require less power. As an example, for the
\Kepler\ response function, a monochromatic laser at $\lambda\sim600$\,nm could
cloak the Earth from \Kepler-like surveys by reaching a peak power of $\sim30$\,MW.
In Figure~\ref{fig:cloak}, we show the expected transit of the Earth as viewed by 
a civilization lying along the ecliptic ($b=0$) using a \Kepler-like bandpass, 
including the effects of limb darkening.

The implications of having to assume a bandpass for the observer are discussed
further in \S\ref{sub:laserreq} and a ``chromatic cloak'' effective at all wavelengths
is introduced in \S\ref{sub:chromatic}.

The effective time that the transmitter needs to be activated at the peak power level
(accounting for the trapezoidal shape of the transit light curve), is given by $\tilde{T}$.
Here, $\tilde{T}$ is the time it takes for a transit to occur, starting from the 
planet's center overlapping with the stellar limb, to exiting the stellar disk 
under the same condition. A reasonable approximation of this duration is given by

\begin{align}
\tilde{T} &= \frac{3^{1/3} P^{1/3} (1-b^2)^{1/2} }{\pi^{2/3} G^{1/3} \rho_{\star}^{1/3} }.
\end{align}

\noindent where $b$ is the impact parameter of the transit (related to the orbital inclination
of the planet's orbit relative to the target star), $\rho_{\star}$ is the mean density
of the parent star and $P$ is the orbital period of the planet to be cloaked.
For a planet receiving comparable insolation to the Earth, the orbital period of this
world would be
	
\begin{align}
P &= \frac{ 2\pi (L_{\star}/L_{\odot})^{3/4} }{ G^{1/2}M_{\star}^{1/2} }\,\mathrm{AU}^{3/2},
\label{eqn:Phab}
\end{align}

\noindent which, with some manipulation, allows us to write that the effective
duration of Earth-like planets would be

\begin{align}
\tilde{T} &= 13\,\mathrm{hrs} \, \sqrt{1-b^2} \, \Big( \frac{ T_{\star} }{ 5777\,\mathrm{K} } \Big) \Big( \frac{ \rho_{\star} }{ \rho_{\odot} } \Big)^{-1/2}.
\end{align}

\noindent Accordingly, the energy usage to cloak each transit is given by

\begin{align}
\Etra =& 760\,\mathrm{MW}\,\mathrm{hrs}\,\sqrt{1-b^2} \Big( \frac{R_P}{R_{\oplus}} \Big)^2 \Big(\frac{\lambda}{600\,\mathrm{nm}}\Big)^2 \Big( \frac{ D_{\mathrm{t}} }{ 10\,\mathrm{m} } \Big)^{-2} \nonumber \\
\qquad& \times \Big( \frac{ T_{\star} }{ 5777\,\mathrm{K} } \Big)^{5} \Big( \frac{ \rho_{\star} }{ \rho_{\odot} } \Big)^{-1/2}.
\end{align}

\noindent The total energy required can be collected over the span of the planet's 
orbital period, $P$, assuming it can be stored, is

\begin{align}
\bar{\Ptra} =&\,\Big(\frac{\Etra}{P}\Big).
\end{align}

\noindent Substituting in the expression for $P$, Equation~(\ref{eqn:Phab}), gives

\begin{align}
\bar{\Ptra} =&\,87\,\mathrm{kW} \sqrt{1-b^2} \,\Big( \frac{R_P}{R_{\oplus}} \Big)^2 \Big(\frac{\lambda}{600\,\mathrm{nm}}\Big)^2 \Big( \frac{ \Dtra }{ 10\,\mathrm{m} } \Big)^{-2} \Big( \frac{ T_{\star} }{ 5777\,\mathrm{K} } \Big)^{2}.
\end{align}

In the picture described, the advanced civilization would continuously generate 87\,kW of 
power over the course of its orbit, storing up the energy and then releasing it with a peak
power of $58$\,MW during the relatively brief transit event. For comparison, the solar
cells on the International Space Station generate 84--120\,kW. However, we point out
cloaking from $N$ targets would of course require $(87 N)$\,kW average power production.
These power requirements are small compared to the GW--TW power requirements
of laser-driven light sailing \citep{guillochon:2015}, meaning that cloaking could be a 
secondary function of laser arrays designed primarily for propulsion.

\subsection{Laser Requirements}
\label{sub:laserreq}

A detailed discussion of the engineering requirements is beyond the scope of this
work, which merely aims to investigate the first-order feasibility of a cloak.
Nevertheless, here we briefly discuss some obvious technical concerns and
suggest they are tractable.

The monochromatic nature of lasers at first seems problematic, but even with a single
beam, the cloak would remain effective at masking planets from \Kepler-like surveys.
Most transit surveys are optimized for Sun-like stars, which peak at 
$\lambda\sim600$\,nm, for reasons largely motivated by the Copernican Principle. 
Rather than use a narrow bandpass filter on \Kepler, the instrument is designed to 
collect as many photons as possible surrounding this key wavelength, in order to
maximize signal-to-noise. For this reason, a monochromatic laser at an optical
wavelength would be still be effective at cloaking a planet from \Kepler-like surveys.
Whilst the cloak would be imperfect since the exact power requirement would depend
upon the sensitivity curve of the transit survey, the transit would be sufficiently
attenuated that the world would either be undetectable or apparently too small
to be habitable. Examples of this are shown in Figure~\ref{fig:cloak}, where we show
how for TESS, CoRoT, MOST and PLATO the residuals would still be below 15\,ppm,
suppressing the transit amplitude by a factor of six.

For our calculations a continuous beam is assumed during the transit event, but the 
same effect could be achieved using short pulses, provided that the interval between 
the pulses is less than the integration time of the transit survey against which one
wishes to cloak. Alternatively, an array of pulsed lasers could be used where the 
relative phasing between them produces near-continuous emission.

For targets closer than 200\,pc, the arriving beam width is less than 2\,AU.
For nearby systems, this leads the requirement of some knowledge of the position of
a planet. If the inclination of the target planet is unknown, then the positional
uncertainty is essentially the semi-major axis, making effective cloaking only possible
through deliberate beam dispersion, increasing the energy requirements. Otherwise,
if the inclination is known, the uncertainty in position is approximately the error in
the semi-major axis, $\Delta a$, indicating that we require $\lesssim$1\% uncertainty
for all systems beyond 1\,pc (using Equation~\ref{eqn:Wrec3}). If the advanced
civilization were to use a Lagrange point for their transit survey, then this would
displace the detector from the assumed target by up to $\tfrac{\sqrt{3}}{2}a$, which using 
Equation~(\ref{eqn:Wrec3}) suggests that cloaking the Earth from advanced civilizations 
closer than 100\,pc would require us to deliberate disperse the beam, somewhat increasing the
energy demands.

In Equation~(\ref{eqn:powerassumption}), it was assumed that the laser beam has a constant
intensity across the beam profile. In reality, the intensity would follow an Airy function
in the far-field limit considered here \citep{guillochon:2015}, meaning that small pointing 
errors much less than that of Equation~(\ref{eqn:pointing}) would lead to imperfect 
cancellation of the transit light curve. Beam shaping using a large number of beams could
potentially solve this problem, which has the extra advantage of relieving the peak power
demand of each laser. Whilst the engineering requirements for such beam shaping
are beyond the scope of this work, we highlight that this problem appears tractable.

If another star system happens to lie behind the target and within the beam,
it would also be subject to the cloaking effect.
Eliminating the transit signal requires careful timing but the transit parallax
effect \citep{scharf:2007} could introduce errors of $\gtrsim1$\,s for targets closer
than 50\,pc and thus the background target could experience imperfect cloaking.
However, such a situation is unlikely given that the angular area of the beam on the sky
is $\pi(\theta/2)^2$, which is $2.4\times10^{-4}\,\mathrm{arcsecond}^2$ for $\Dtra=10$\,m and
$\lambda=600$\,nm. For comparison, the URAT1 survey \citep{zacharias:2015} reveals that
even the highest density regions of the sky (near the galactic plane) have 
$\lesssim 8 \times 10^{-3}$ stars per squared arcsecond down to $R<18.5$, suggesting the
maximum number of stars within a beam is $\sim30$. The fraction of stars which have inhabited
worlds harboring technological civilizations is less than 1 in 30 for even the most optimistic
estimates and thus we consider this to be a highly unlikely situation.

\begin{figure}
\begin{center}
\includegraphics[width=8.4 cm]{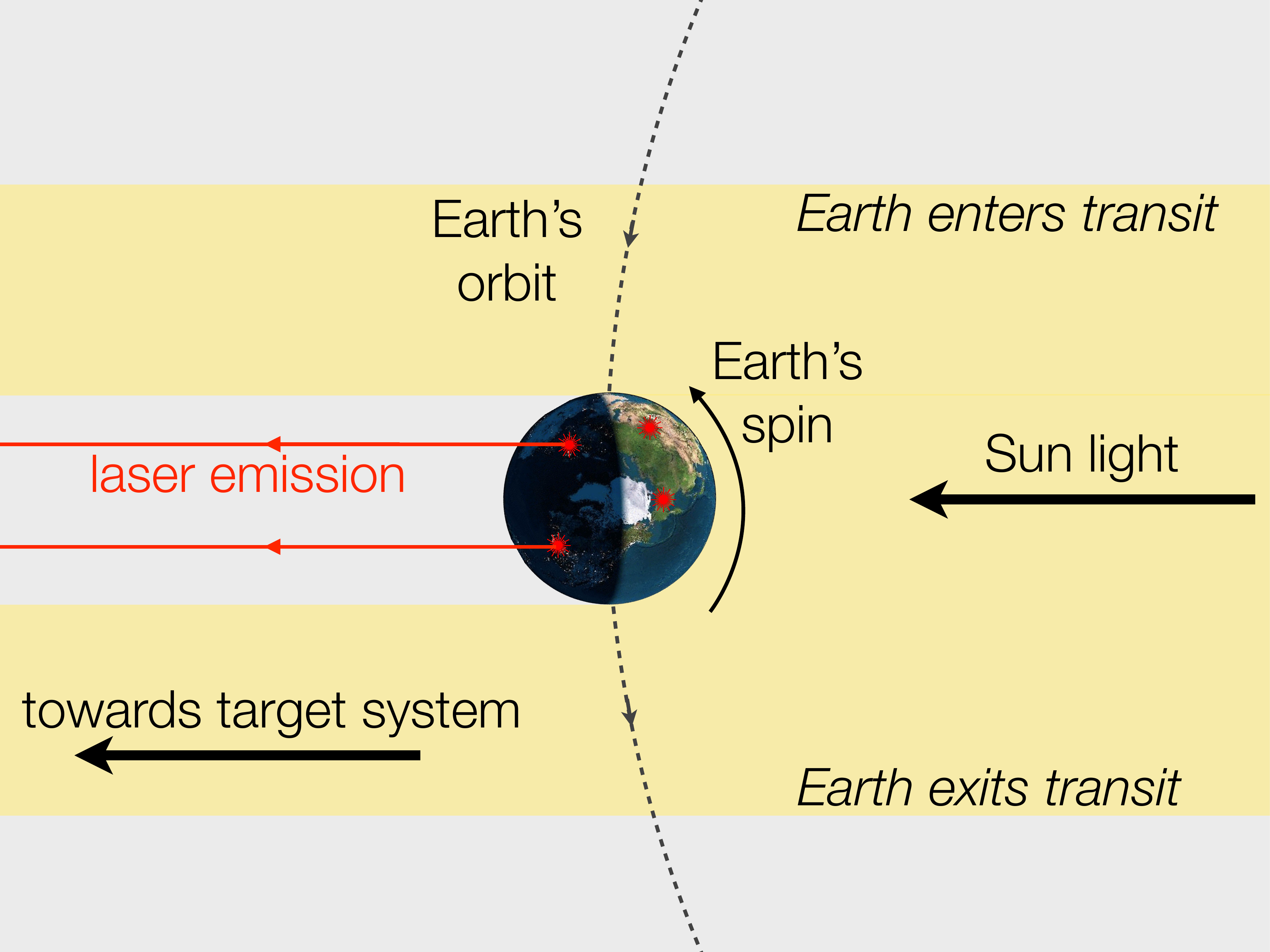}
\caption{
Illustration (not to scale) depicting the Earth's revolution and rotation as
it passes through the transit cone of a distant target system. For ground-based
emission, numerous sites would be required to compensate for the Earth's rotation,
as shown.
}
\label{fig:geometry}
\end{center}
\end{figure}

The selection of a suitable location for the laser array bears mentioning. Clearly to cloak 
a transit the lasers must be beamed at the antisolar point, but the rotation of the planet
complicates the selection of any geographic location for the array. More than one site would 
be needed, as a single location would be in daylight for a significant fraction of the time 
and would therefore be unable to shine at the antisolar point (see Figure~\ref{fig:geometry}). 
The arrays would need to pivot to adjust for azimuthal and altitudinal changes continuously 
like any tracking telescope, and the lasers would suffer from absorption in Earth's atmosphere. 
The absorption would also change continuously as a function of atmospheric path length. 

One way to eliminate these difficulties would be to place the laser array in 
space at the third or fourth Lagrange point (L$_4$/L$_5$). Of course, locating
the array in space would have its own challenges: it would require a dedicated
mission for installation, sufficient power would have to be generated by solar 
panels alone, and maintenance would be non-trivial if possible at all. We omit here a 
discussion on the relative merits of a terrestrial versus a space-based solution. 
It is worth pointing out, however, that placing the laser array at L$_4$/L$_5$ means 
that any planet, not just the inhabited world, could be cloaked or used for 
broadcasting (see \S\ref{sec:broadcast}). An extraterrestrial civilization might 
therefore opt to utilize the transit of the inner most planet in their system for
this purpose (e.g. a hot-Jupiter), as it would be particularly efficient for 
signaling given the short orbital period.

\subsection{Chromatic Cloak}
\label{sub:chromatic}

A chromatic cloak (a cloak effective at all wavelengths) could be achieved using
a large number of beams, rather than one. This would have the additional advantage
of relieving the peak power requirement per beam. Since the actual origin of the
lasers is unresolved from the parent star, using multiple lasers at different
wavelengths is indistinguishable from a single source at multiple wavelengths.
Each of these beams would emit at a different wavelength, with the power tuned to
the spectral radiance function of the parent star.

To calculate the total chromatic power required to cloak a planet, we modify
Equation~(\ref{eqn:Pt}) to define the spectral power, $\tilde{\mathcal{P}}$,
of the transmitter as

\begin{align}
\tilde{\mathcal{P}}_{\mathrm{t}} (\lambda)\,\mathrm{d}\lambda= \Bigg( \frac{R_P^2 \lambda^2}{2 D_t^2 R_{\star}^2} \Bigg) \tilde{\mathcal{P}}_{\star,\mathrm{t}}(\lambda)\,\mathrm{d}\lambda,
\label{eqn:Pt_spec}
\end{align}

\noindent where for blackbody-like stellar radiation, we have

\begin{align}
\tilde{\mathcal{P}}_{\star,\mathrm{t}}(\lambda)\,\mathrm{d}\lambda &= \big(4\pi R_{\star}^2\big) \Bigg(\frac{2 h c^2}{\lambda^5} \frac{\pi}{e^{\frac{h c}{\lambda k_B T_{\star}}}-1}\Bigg)\,\mathrm{d}\lambda.
\label{eqn:Pstar_spec}
\end{align}

\noindent Here, $\tilde{\mathcal{P}}_{\mathrm{t}} (\lambda)$ and 
$\tilde{\mathcal{P}}_{\star,\mathrm{t}}(\lambda)$ are
the wavelength-specific transmission power and fractional stellar luminosities, respectively,
such that the spectrally integrated power is

\begin{align}
\mathcal{P} &= \int_{\lambda=0}^{\infty} \tilde{\mathcal{P}}(\lambda)\,\mathrm{d}\lambda.
\end{align}

Accordingly, the total power required to the transmitter array of a chromatic cloak is

\begin{align}
\mathcal{P}_{\mathrm{t}} =&\,\Bigg( \frac{4 \pi^2 h c^2 R_P^2}{D_t^2} \Bigg) \int_{\lambda=0}^{\infty}
\frac{1}{\lambda^3} \frac{1}{e^{\frac{h c}{\lambda k_B T_{\star}}}-1}\,\mathrm{d}\lambda,
\label{eqn:integrand}
\end{align}

\noindent for which a closed-form expression is obtainable as

\begin{align}
\mathcal{P}_{\mathrm{t}} =& \sigma_C \Bigg(\frac{R_P^2 T_{\star}^2}{D^2}\Bigg),
\label{eqn:Pt_int}
\end{align}

\noindent where we define $\sigma_C = \tfrac{2}{3} \pi^4 k_B^2 h^{-1} = 1.86818\times10^{-11}$\,J\,s\,K$^{-2}$.
Evaluating Equation~(\ref{eqn:Pt_int}) in canonicalized units, one may write

\begin{align}
\mathcal{P}_{\mathrm{t}} (\lambda) =& 254\,\mathrm{MW}\,\Bigg( \frac{R_P}{R_{\oplus}} \Bigg)^2
\Bigg( \frac{T_{\star}}{5777\,\mathrm{K}} \Bigg)^2
\Bigg( \frac{D}{10\,\mathrm{m}} \Bigg)^{-2}.
\label{eqn:Pt_int2}
\end{align}

Clearly an infinite series of monochromatic lasers spaced at infinitesimal
wavelength intervals is not practical and thus the effectiveness of any
chromatic cloak will depend upon the wavelength resolution. If the the observer
is taking spectroscopic transit observations with a spectral resolution power of $R$,
then in the wavelength range of $\lambda_1=\lambda_{\mathrm{min}}$ to
$\lambda_{N}=\lambda_{\mathrm{max}}$, one would require $N$ monochromatic lasers, 
where each laser would operate at a wavelength of

\begin{align}
\lambda_i &= \lambda_{\mathrm{min}} \Bigg( \frac{1+R^{-1}/2}{1-R^{-1}/2} \Bigg)^{i-1},
\label{eqn:lambda_i}
\end{align}

\noindent and requiring a total number of lasers equal to

\begin{align}
N &= 1 + \frac{ \log(\lambda_{\mathrm{max}}/\lambda_{\mathrm{min}}) }{ \log\big( 1 + \frac{2}{2R-1} \big) }.
\end{align}

\noindent For $R\gg1$, this is well approximated by

\begin{align}
N &\simeq R \log(\lambda_{\mathrm{max}}/\lambda_{\mathrm{min}}).
\end{align}

\noindent For example, a chromatic cloak able to defeat the $R=2700$ integral field
NIRSpec instrument planned for JWST from 0.6--5\,$\mu$m would require approximately
6000 monochromatic lasers in the array. The power of each laser, $p_{\mathrm{t},i}$ may be
estimated by taking the limit of Equation~(\ref{eqn:integrand}) with infinitesimal
integral limits, giving

\begin{align}
p_{\mathrm{t},i} =&\,\Bigg( \frac{4 \pi^2 h c^2 R_P^2}{R D_t^2} \Bigg)
\Bigg( \frac{1}{\lambda_i^2} \frac{1}{e^{\frac{h c}{\lambda_i k_B T_{\star}}}-1} \Bigg),
\label{eqn:spectralpower}
\end{align}

\noindent where the total power fed to the transmitter array, $\mathcal{P}_t$,
is equal to the sum of these lasers such that $\mathcal{P}_t = 
\sum_{i=1}^N p_{\mathrm{t},i}$. Equation~(\ref{eqn:spectralpower}) can be shown 
to be maximized at a wavelength of

\begin{align}
\lambda_{p,\mathrm{max}} &= 1.563\,\mu\mathrm{m} \Bigg( \frac{T_{\star}}{5777\,\mathrm{K}} \Bigg)^{-1}.
\end{align}

\noindent Evaluating the power per wavelength bin function at this maximum for a Sun-like
$T_{\star}$ allows us to write

\begin{align}
p_{\mathrm{t},i} \leq&\,100\,\mathrm{kW} \Bigg( \frac{R}{1000} \Bigg)^{-1}.
\end{align}

In summary then, a chromatic cloak may be achieved by an array of lasers
emitting at controlled and regularly spaced wavelength intervals each with
a unique peak power given by Equation~(\ref{eqn:spectralpower}). Emerging
technologies, such as tunable lasers which can be swept across the spectrum
at high speed, so-called supercontinuum lasers (e.g. see \citealt{cumberland:2008}), 
may allow for direct simulation of any spectrum desired, thereby negating some of 
the technical issues discussed above. Whilst a detailed technology
feasibility study is beyond the scope of this paper, we see no reason why
advanced civilizations could not build such devices given that we are on the
cusp of such abilities ourselves.

\section{Cloaking Biosignatures}
\label{sec:biocloak}

\subsection{Concept}

As a more energy efficient alternative to cloaking the presence of the planet, it 
may be desirable to cloak only certain atmospheric features, in particular, atmospheric
constituents that may indicate the presence of life (so-called ``biosignatures'' or 
industrial pollutants; \citealt{kaltenegger:2002,lin:2014}) or otherwise favorable
conditions for life (such as the presence of water vapor). We refer to this as a
``biocloak'' in what follows.

It is straightforward to use a chromatic laser array to cancel
out the absorption features in the planet's transmission spectrum, assuming laser 
emission can be produced at any desired wavelength. Indeed, the presence of an atmosphere 
could be cloaked altogether if the effective height changes of the planet as a function 
of wavelength are canceled out by lasers. The planet might then resemble a dead world 
totally devoid of any atmosphere and appear almost certainly hostile to life. Not only 
would this approach require a significantly smaller power output, it would also have the 
benefit of producing self-consistent observations insomuch as the presence of the planet 
might still be inferred by other means (i.e. through radial velocity analysis). Since 
we do not here present a means for disguising a radial velocity signature, the presence 
of such a signature without a corresponding transit could raise suspicions that the 
planet's transit was deliberately cloaked, thus inadvertently indicating the presence 
of a technologically advanced civilization.

\subsection{A Biocloak for the Earth}

Since the atmospheres of inhabited exoplanets may be diverse and quite exotic
(e.g. see \citealt{seager:2013}) we consider here the case of cloaking the Earth's 
atmosphere as a concrete example. To estimate the power output required, we employ the 
\citet{kaltenegger:2009} model of terrestrial effective height as a function of wavelength
in the 0-20\,$\mu$m range. This region shows a number of important absorption features in 
the Earth's atmosphere, including water and several molecules typically considered 
biomarkers when found in extreme thermodynamic disequilibrium (e.g. O$_2$, O$_3$, and 
CH$_4$). From here, we can calculate the change in flux associated with these absorption 
features by

\begin{align}
\frac{\Delta F(\lambda)}{F(\lambda)} = \left( \frac{R_P+h(\lambda)}{R_{\star}} \right)^2 - \left( \frac{R_P}{R_{\star}} \right)^2
\end{align}

\noindent where $h$ is the effective height. To calculate the total 
chromatic power required to cloak the atmosphere, we modify Equation~(\ref{eqn:Pt_spec}) 
such that

\begin{align} \label{Pt_atmos}
\tilde{\mathcal{P}}_{\mathrm{t}} (\lambda)\,\mathrm{d}\lambda = \left( \frac{\lambda ^2 [(R_P+h(\lambda))^2-R_P^2]}{2D_t^2 R_{\star}^2} \right) \tilde{\mathcal{P}}_{\star,\mathrm{t}}(\lambda)\,\mathrm{d}\lambda
\end{align}

Taking $D_t$ = 10\,m and using a dynamic spacing between laser wavelengths as determined 
by Equation~(\ref{eqn:lambda_i}), the total power required for atmospheric cloaking 
over the 20\,$\mu$m range is then just the sum across all wavelengths. Clearly lasers spaced 
at larger wavelength intervals will have to compensate with more power, and will result in a 
more unnatural spectral appearance if the observing instrument has sufficiently high 
resolution. It is therefore desirable to minimize the spacing of the laser wavelengths as 
much as is feasible.

For the Earth and in this wavelength range, we find that the integrated sum of the peak powers
across all wavelengths is $\sim$1\,MW. In fact, if we subtract the minimum effective height
($\sim$6\,km) from the atmospheric thickness and add it to the planet's radius this number can
be reduced to $\sim$607\,kW. Chromatic atmospheric cloaking in the 0-20\,$\mu$m range then requires only
$\sim$2\% the power required for broadband transit cloaking (up to 30\,MW at 600\,nm).

Selective cloaking of certain atmospheric features will reduce power requirements further still. 
In principle, it could be possible to mimic a wide range of atmospheric compositions by 
artificially adjusting absorption features to any desired depth. We consider here the case of cloaking 
the absorption features of molecular oxygen and ozone in the 0-20\,$\mu$m range, which we
label as a ``biocloak''. For this calculation we derive the effective height due to these molecules 
from the relative transmission spectrum published in \cite{kaltenegger:2009}. With values for 
$h(\lambda)$ in hand we repeat the process from before to derive the power requirements. We find a 
total power requirement $\sim$160\,kW to mask O$_2$ and O$_3$. In this case we cannot subtract any 
baseline as there are some wavelengths for which $h$ is zero.

\begin{figure}
\begin{center}
\includegraphics[width=8.4 cm]{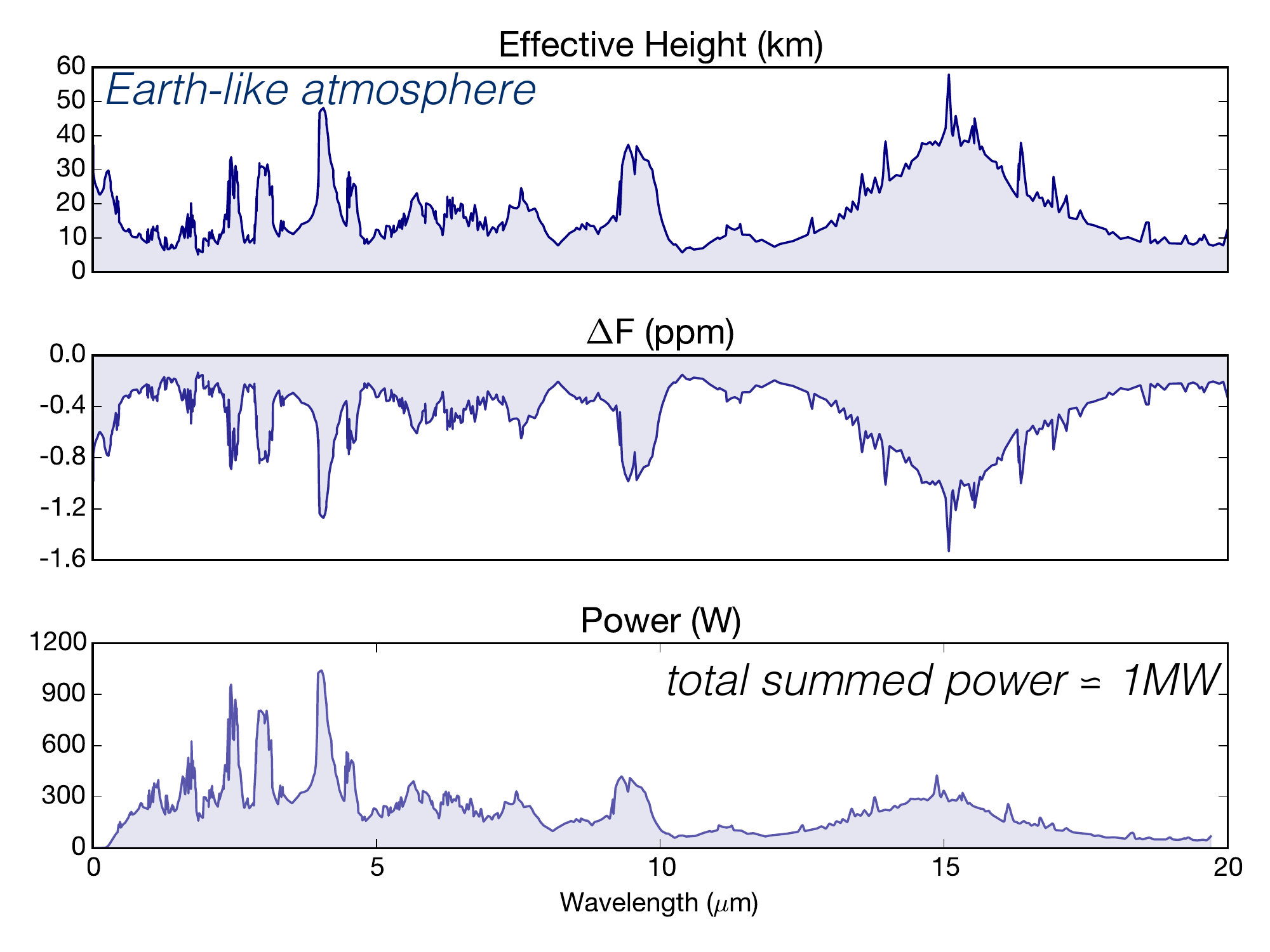}
\caption{
\textit{Top:} Effective height of the Earth's atmosphere in km, taken from \citep{kaltenegger:2009}. 
\textit{Middle:} The reduction in flux associated with these absorption features in parts per million.
\textit{Bottom:} Power requirement $\mathcal{P}_{\mathrm{t}} (\lambda)$ in W, assuming a solar 
blackbody with $T_{eff} = $ 5777\,K and chromatic lasers spaced dynamically according to 
Equation~(\ref{eqn:lambda_i}) and setting $R=1000$.
}
\end{center}
\end{figure}

\begin{figure}
\begin{center}
\includegraphics[width=8.4 cm]{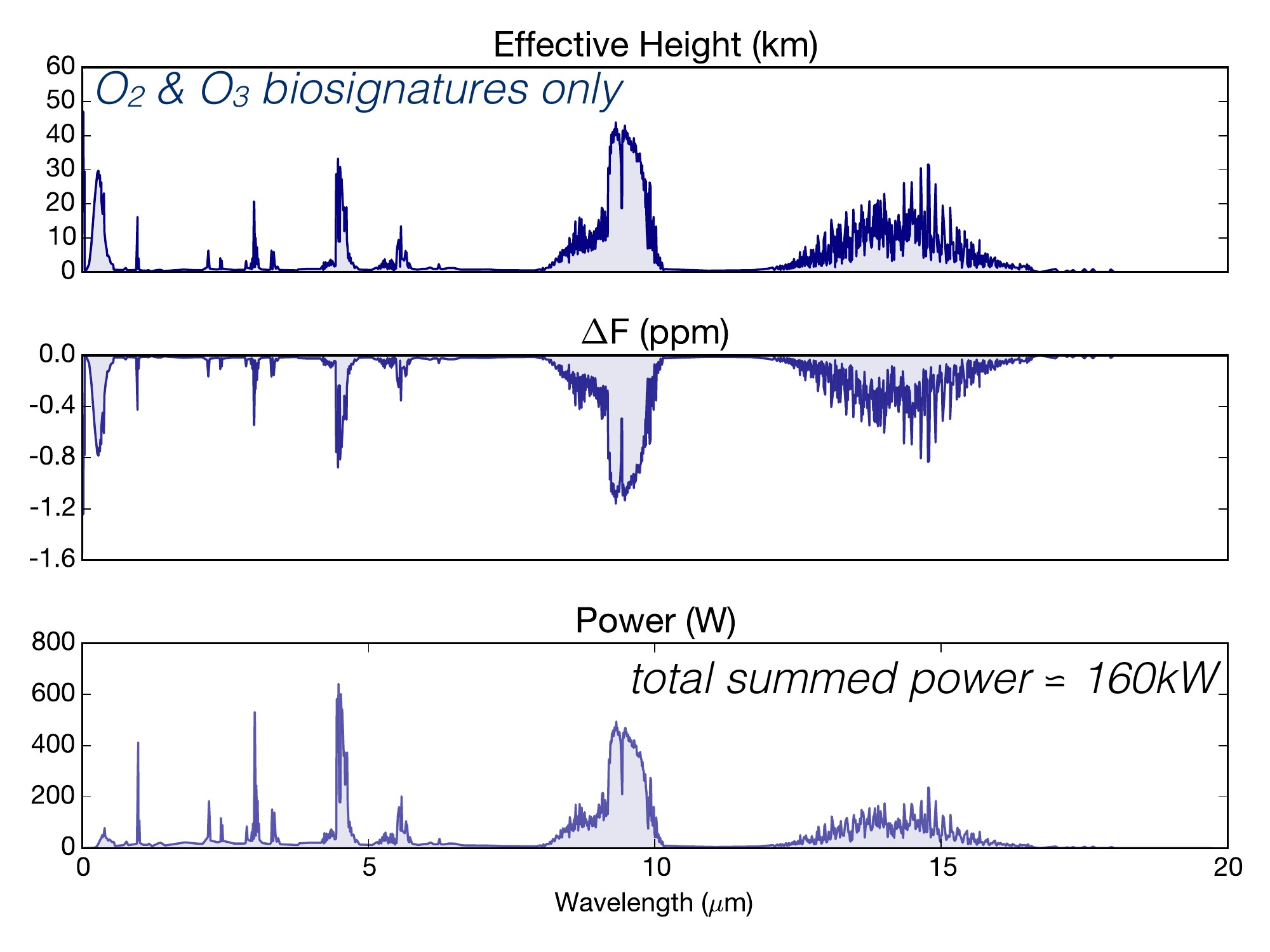}
\caption{
\textit{Top:} Effective height of the Earth's atmosphere in km due solely to molecular oxygen and 
ozone, derived from \citep{kaltenegger:2009}. \textit{Middle:} The reduction in flux associated with 
these absorption features in parts per million. \textit{Bottom:} Power requirement 
$\mathcal{P}_{\mathrm{t}} (\lambda)$ in W, assuming a solar blackbody with $T_{eff} = $ 5777\,K and 
chromatic lasers spaced dynamically according to Equation~(\ref{eqn:lambda_i}) and setting $R=1000$.
}
\end{center}
\end{figure}

\section{Broadcasting A Planet}
\label{sec:broadcast}

\subsection{Concept}

Finally, we introduce a third type of laser emission which performs the 
opposite function of cloaking, namely deliberate broadcasting. Indeed, as 
outlined in \S\ref{sec:intro}, this formed the original inspiration behind 
our idea - to find a more practical approach to creating artificial transits
versus the \citet{arnold:2005} mega-structure concept.

Whilst any number of artificial transit profiles can be created with 
lasers, one ideally seeks a profile which is both energy efficient and 
unambiguously artificial. Producing upward spikes in-transit might seem like an
obvious suggestion, but star spot crossings produce these forms with complex
and information rich signatures (e.g. see \citealt{beky:2014}). Here,
we argue that cloaking the ingress/egress of a transit, but leaving the main
transit undistorted, would be a highly effective strategy since no known
natural phenomenon is likely to produce such an effect.

The duration of the ingress/egress ($T_{12}$) is a proxy for the impact 
parameter, $b$, of the transit. The shortest ingress/egress time possible 
occurs for equatorial transits, when $b=0$. By measuring both the 
first-to-fourth and second-to-third contact durations of a transit,
one may solve uniquely for both $b$ and the mean stellar density, $\rho_{\star}$
\citep{seager:2003}.

\begin{figure}
\begin{center}
\includegraphics[width=8.4 cm]{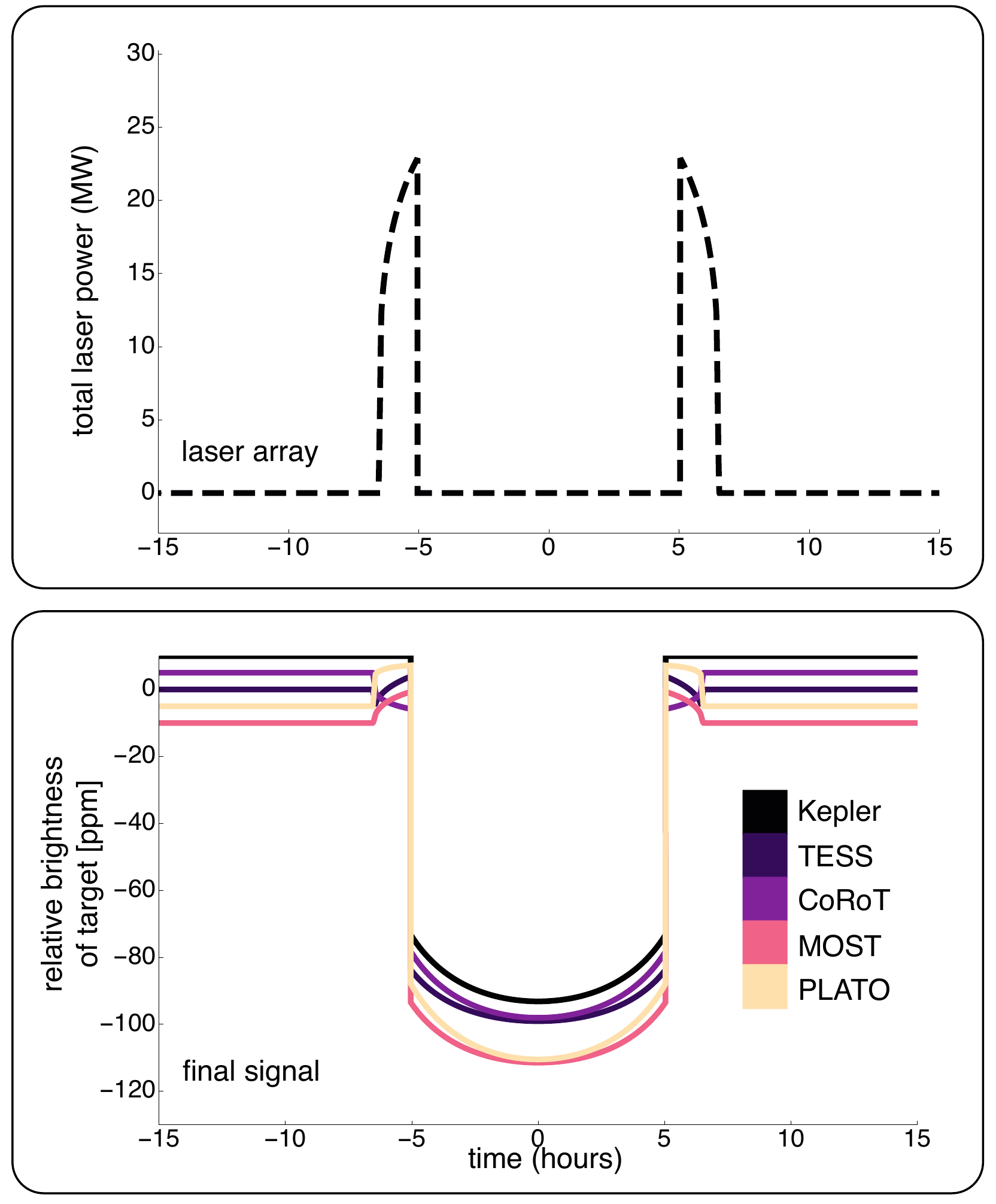}
\caption{
Same as Figure~\ref{fig:cloak}, except for the case of broadcasting rather
than cloaking.
\textit{Top:} The power profile of a laser array designed to broadcast the 
Earth. An array of lasers producing a peak power of $\sim20$\,MW for
approximately 15\,minutes nullifies the transit ingress and egress. 
\textit{Bottom:} The resulting light curves as viewed by different broadband
optical photometers. The observed impact parameter would be complex infinity,
for which a normal light curve fit would be unable to explain and thus
indicating the presence of artificial transit manipulation.
}
\label{fig:broadcast}
\end{center}
\end{figure}

For any transiting planet, one should always expect $T_{12}\geq\lim_{b\to0}T_{12}$ 
and certainly an instantaneous ingress/egress cannot be produced by any physical
phenomenon presently known to us. This geometric boundary condition provides an 
avenue to broadcast a clear artificial signature. Instead of cloaking the full 
transit then, cloaking just the ingress/egress components would create an 
artificial transit, as shown in Figure~\ref{fig:broadcast}.

The depth of the transit is unaltered, meaning that the observer is still able
to accurately infer $(R_P/R_{\star})$. Assuming the observer is able to independently
measure a near-zero eccentricity and the mean stellar density, the inferred
impact parameter is given by Equation~(7) of \citet{seager:2003}

\begin{align}
b &= \sqrt{\frac{(1-p)^2 - f^2 (1+p)^2}{1-f^2}},
\end{align}

where $f$ is approximately the ratio $T_{23}/T_{14}$. Accordingly, by cloaking
the ingress/egress the observer finds $f\to1$ leading to an imaginary impact
parameter for all $f>(1-p)/(1+p)$, which is 0.98 for the Earth-Sun transit.
Therefore, artificial transits should be revealed by their imaginary impact parameters.

Broadcasting has the advantage that one does not need to be concerned with perfect
cloaking across all wavelengths any longer. For example, the laser emission
could be monochromatic here, which would be initially identified
as an artificial transit by a \Kepler-like survey and then follow-up spectroscopy
would find clear evidence for laser emission, further supporting an artificial
origin. Moreover, information could be beamed along the laser for the purposes of
communication. In this sense, the transits provide both an initial way of signaling
our presence, like a beacon, and also a suitable short-window in which to attempt
active communication.

\subsection{Energy Requirements}

For completion, we estimate the energy requirements to accomplish broadcasting,
as was done earlier for cloaking. The peak power emission required will be of
the same order as that needed for cloaking, with the only difference being
that the emission time can be reduced from many hours to many minutes, thereby
reducing the total energy requirement. Assuming a circular orbit for simplicity,
the ingress/egress time can be approximated to

\begin{align}
T_{12} &= \frac{ 0.12\,\mathrm{hours} }{ \sqrt{1-b^2} } \Big( \frac{ R_P }{ R_{\oplus} } \Big) \Big( \frac{ L_{\star} }{ L_{\odot} } \Big)^{1/4} \Big( \frac{ M_{\star} }{ M_{\odot} } \Big)^{-1/6} \Big( \frac{ R_{\star} }{ R_{\odot} } \Big)^{-1} \Big( \frac{ \rho_{\star} }{ \rho_{\odot} } \Big)^{-1/3},
\end{align}

\noindent which can be further simplified to

\begin{align}
T_{12} &= \frac{ 0.12\,\mathrm{hours}} { \sqrt{1-b^2} } \Big( \frac{ R_P }{ R_{\oplus} } \Big) \Big( \frac{ L_{\star} }{ L_{\odot} } \Big)^{1/4} \Big( \frac{ M_{\star} }{ M_{\odot} } \Big)^{-1/2}.
\end{align}

\noindent Accordingly, we estimate that the total energy needed to cloak
the ingress and egress for a single transit event would be

\begin{align}
\mathcal{E}_{\mathrm{t}} =& \frac{ 6.9\,\mathrm{MW}\,\mathrm{hours} }{ \sqrt{1-b^2} }\,\Big( \frac{R_P}{R_{\oplus}} \Big)^3 \Big(\frac{\lambda}{600\,\mathrm{nm}}\Big)^2 \Big( \frac{ D_{\mathrm{t}} }{ 10\,\mathrm{m} } \Big)^{-2} \nonumber\\
\qquad& \times \Big( \frac{ R_{\star} }{ R_{\odot} } \Big)^{-2} \Big( \frac{ L_{\star} }{ L_{\odot} } \Big)^{5/4} \Big( \frac{ M_{\star} }{ M_{\odot} } \Big)^{-1/2},
\end{align}

\noindent which per orbital period of an Earth-like insolation planet equates to an averaged power of

\begin{align}
\bar{\mathcal{P}}_{\mathrm{t}} =& \frac{ 6.9\,\mathrm{MW}\,\mathrm{hours}\,\mathrm{year}^{-1} }{ \sqrt{1-b^2} }\,\Big( \frac{R_P}{R_{\oplus}} \Big)^3 \Big(\frac{\lambda}{600\,\mathrm{nm}}\Big)^2 \Big( \frac{ D_{\mathrm{t}} }{ 10\,\mathrm{m} } \Big)^{-2} \nonumber\\
\qquad& \times \Big( \frac{ R_{\star} }{ R_{\odot} } \Big)^{-2} \Big( \frac{ L_{\star} }{ L_{\odot} } \Big)^{1/2}.
\end{align}

\noindent To compare the requirements for different host stars, we may adopt
$L_{\star} \propto M_{\star}^4$ and $M_{\star} \propto R_{\star}$ to re-write the
above as

\begin{align}
\bar{\mathcal{P}}_{\mathrm{t}} &= \frac{ 720\,\mathrm{W} }{ \sqrt{1-b^2} }\,\Big( \frac{R_P}{R_{\oplus}} \Big)^3 \Big(\frac{\lambda}{600\,\mathrm{nm}}\Big)^2 \Big( \frac{ D_{\mathrm{t}} }{ 10\,\mathrm{m} } \Big)^{-2} \Big( \frac{ M_{\star} }{ M_{\odot} } \Big)^{2}.
\end{align}

\noindent It therefore requires two orders of magnitude less energy (or orbit averaged power) to
broadcast a planet than cloak it, due to the fact the ingress/egress is two
orders of magnitude shorter in duration.

\section{Discussion}
\label{sec:discussion}

In this work, we suggest that directed laser emission may be used to
either broadcast or cloak the presence of the Earth to extraterrestrial
transit surveys. Motivated by earlier proposals of searching for 
mega-structures designed to create artificial transits \citep{arnold:2005},
we argue that artificial transits could be produced with presently 
available technology. On this basis, a search for artificial transit
signatures, similar to those described in this work, may be a worthwhile
search strategy for non-radio SETI.

We estimate that cloaking the Earth from a broadband \Kepler-like survey could
be achieved with an array of monochromatic optical lasers emitting a peak
power of $\sim30$\,MW. In this scheme, one would target a known Earth-analog
residing near the ecliptic (and thus can see the Earth transit) and emit
for $\sim10$\,hours once per year. A $\mathcal{O}[10\,\mathrm{m}^2]$ solar cell array
would be sufficient to gather this energy over the course of a year, meaning
that the laser array could be feasibly placed in space. We estimate that a chromatic
cloak (effective at all wavelengths) would be an order-of-magnitude more costly
energetically.

Deliberate broadcasting relaxes some of the technical challenges of chromatic
cloaking and requires three orders of magnitude less energy. We argue that cloaking 
of the transit ingress/egress alone would present a clear artificial signature, 
one which we could begin surveying archival \Kepler\ data for immediately. The time 
of transit provides a natural communication window, analogous to water hole in radio
SETI \citep{oliver:1979}. This line of thought can be extended to the time of
inferior conjunction for non-transiting planets too. Advantageously, our method allows one
to simultaneously alert an observer to an artificial source (serving the same function
as radio SETI beacons; \citealt{benford:2008}) and transit large volumes of data along
the beams.

Whilst it is possible to actively begin searching for broadcasted signals in, say,
archival \Kepler\ photometry, cloaked exoplanets obviously present a challenge by
their very design. Transits are not the only method to discover planets and thus
a truly xenophobic civilization may conclude that even a perfect and chromatic transit
cloak would be ultimately defeated by observation of the planet using radial
velocities. In this sense, the biocloak is perhaps the most effective
strategy since certainly the transit and radial velocity measurements would appear
compatible. However, even here, direct imaging would reveal a strong discrepancy
in terms of the atmospheric interpretation and thus overcome the cloak.

For these reasons, perhaps the most effective use of laser enabled transit distortion
would be for broadcasting rather than cloaking. In such a case, it may be more
effective to construct the laser array not in the vicinity of L$_4$/L$_5$ for their 
own home planet, but at L$_4$/L$_5$ for the shortest period planet in their system. 
Therefore, in the Solar System we would distort the transit of Mercury. This provides 
a higher duty cycle of distorted events, facilitating their detection. We therefore 
suggest that any survey in archival data should not be limited to rocky planets in the 
habitable-zone of their host star.
	
\section*{Acknowledgments}

We thanks Jason Wright and the Cool Worlds Lab members for useful
discussions in preparing this paper. Special thanks to the anonymous
reviewer for their helpful comments.

\bsp
\label{lastpage}
\end{document}